\documentstyle[aps, manuscript]{revtex}
\tightenlines
\draft
\begin{document}
\title{Noncompact Kaluza-Klein theory and inflationary cosmology:
a complete formalism }
\author{$^1$Jos\'e Edgar Madriz Aguilar\footnote{
E-mail address: edgar@itzel.ifm.umich.mx}
and $^{2}$Mauricio Bellini\footnote{
E-mail address: mbellini@mdp.edu.ar}}
\address{$^1$Instituto de F\'{\i}sica y Matem\'aticas,
AP: 2-82, (58040) Universidad Michoacana de San Nicol\'as de Hidalgo,
Morelia, Michoac\'an, M\'exico.\\
$^2$ Consejo Nacional de Investigaciones Cient\'{\i}ficas y
T\'ecnicas (CONICET) \\
and\\
Departamento de F\'{\i}sica, Facultad de Ciencias Exactas y Naturales,
Universidad Nacional de Mar del Plata,
Funes 3350, (7600) Mar del Plata, Argentina.}

\vskip .5cm
\maketitle
\begin{abstract}
A formalization of the recently introduced formalism for inflation
is deve\-loped from a noncompact Kaluza-Klein theory.
In particular, the case for a single scalar field inflationary model is
studied.
We obtain that the scalar potential,
which assume different representations in different
frames, has a geometrical origin.
\end{abstract}
\vskip .2cm                             
\noindent
Pacs numbers: 04.20.Jb, 11.10.kk, 98.80.Cq \\
\vskip 1cm
In the last years, many people has worked in extra dimensions\cite{*}.
As has been emphasized, Standard Model matter can propagate a large
distance in extra dimensions without conflict with observations if
Standard Model is confined to a $(3+1)$-dimensional subspace, or
``3-brane'', in the higher dimensions\cite{a}. It should be
possible if the four familiar dimensions where dependent of coordinate
in the extra dimensions. In some works on non compact extra
dimensions the authors
studied trapping of matter fields to be effectively four
dimensional (4D)\cite{b} or studied finite-volume but topologically
non compact extra dimensions\cite{a1}.

A very important question in theoretical physics consists to
provide a good geometrical description of matter using only
one extra coordinate (say $\psi$).
The explanation of this issue in the framework of the early
universe, in particular for inflationary theory\cite{Guth},
should be of great importance in cosmology.
In this letter,
we are aimed to study this topic using
the Kaluza-Klein formalism where
the fifth coordinate is non compact.
In this framework should be interesting
to explain the origin of an effective
four dimensional (4D) scalar potential
$V(\varphi)$ which could be originated from a
5D apparent vacuum.
The idea that matter in four dimensions (4D) can be explained from
a 5D Ricci-flat ($R_{AB}=0$) Riemannian manifold is a consequence of the
Campbell's theorem. It says that any analytic $N$-dimensional Riemannian
manifold can be locally embedded in a $(N+1)$-dimensional Ricci-flat
manifold. This is of great importance for establishing the generality of the
proposal that 4D field equations with sources can be locally embedded
in 5D field equations without sources\cite{wesson}.
For apparent vacuum we understand
a 5D flat metric and a 5D Lagrangian for a neutral scalar field,
where the 5D dynamics is only kinetic,
that is, the 5D potential in the 5D Lagrangian do not exists.
In other words, we shall consider an 5D apparent vacuum for
scalar fields without sources or interactions.

We consider the 5D metric, recently introduced by Ledesma and
Bellini (LB)\cite{PLB}
\begin{equation}\label{6}
dS^2 = \psi^2 dN^2 - \psi^2 e^{2N} dr^2 - d\psi^2,
\end{equation}
where the parameters ($N$,$r$)
are dimensionless and the fifth coordinate
$\psi $ has spatial unities.
The metric (\ref{6}) describes a
flat 5D manifold in apparent vacuum ($G_{AB}=0$).
To describe neutral matter in a 5D geometrical vacuum
(\ref{6}) we can consider
the Lagrangian
\begin{equation}\label{1}
^{(5)}{\rm L}(\varphi,\varphi_{,A}) =
-\sqrt{\left|\frac{^{(5)}
g}{^{(5)}g_0}\right|} \  ^{(5)}{\cal L}(\varphi,\varphi_{,A}),
\end{equation}
where $^{(5)}g=\psi^8 e^{6N}$ is the determinant of the 5D metric tensor with
components $g_{AB}$ ($A,B$ take the values $0,1,2,3,4$) and $^{(5)}g_0=\psi^8_0 e^{6N_0}$
is a constant of dimensionalization determined
by $^{(5)}g$ evaluated with the initial conditions of the
system: $\psi=\psi_0$ and $N=N_0$. We shall consider $N_0=0$, so that
$^{(5)}g_0=\psi^8_0$.
Since the 5D metric (\ref{6}) describes a manifold in apparent
vacuum, the density Lagrangian
${\cal L}$ in (\ref{1}) must to be
\begin{equation}\label{1'}
^{(5)}{\cal L}(\varphi,\varphi_{,A}) = 
\frac{1}{2} g^{AB} \varphi_{,A} \varphi_{,B},
\end{equation}
which describes a free scalar field because there is no interaction:
$V[\varphi(N,r,\psi)]=0$.
Taking into account
the metric (\ref{6}) and the Lagrangian (\ref{1}), we
obtain the equation of motion for $\varphi$
\begin{equation}\label{df}
\left(2\psi \frac{\partial\psi}{\partial N}+ 3 \psi^2 \right)
\frac{\partial\varphi}{\partial N}
+\psi^2 \frac{\partial^2\varphi}{\partial N^2} - \psi^2 e^{-2N} \nabla^2_r\varphi
-4\psi^3 \frac{\partial\varphi}{\partial\psi} - 3\psi^4 \frac{\partial N}{
\partial\psi} \frac{\partial\varphi}{\partial\psi} - \psi^4 \frac{\partial^2
\varphi}{\partial\psi^2} =0,
\end{equation}
where ${\partial N \over \partial\psi}$ is zero because the coordinates
$(N,\vec{r},\psi)$ are independents.

Now, as in earlier works\cite{PLB,MB},
we can consider the 3D comoving frame $dr=0$.
Taking the metric (\ref{6}) with $U^r=0$,
the geodesic dynamics ${dU^C \over dS}=-\Gamma^C_{AB} U^A U^B$
with $g_{AB} U^A U^B=1$, give us the following velocities $U^A$:
\begin{equation}\label{vel}
U^{\psi} = - \frac{1}{\sqrt{u^2(N)-1}}, \qquad U^{r}=0, \qquad
U^N=\frac{u(N)}{\psi\sqrt{u^2(N)-1}},
\end{equation}
for $S(N)=-N$  and $u(N)={\rm coth}(N)$.
In this representation ${d\psi \over dN}=\psi/u(N)$.
Thus the fifth coordinate evolves as
\begin{equation}\label{psi}
\psi(N) = \psi_0 e^{\int dN/u(N)}.
\end{equation}
Here, $\psi_0$ is a constant of integration that has spatial unities.
From the mathematical point of view, we are taking a foliation
of the 5D metric (\ref{6}) with $r$ constant.
Hence, to describe
the metric in physical coordinates we can make the following
transformations:
\begin{equation}\label{*}
t = \int \psi(N) dN, \qquad R=r\psi, \qquad L= \psi(N) \  e^{-\int dN/u(N)},
\end{equation}
such that for $\psi(t)=1/H_c(t)$ (i.e., for  $u(N)=-{H_c \over dH_c/dN} >0$),
we obtain the resulting 5D metric
\begin{equation}\label{fr}
dS^2 = dt^2 - e^{2\int H_c(t) dt} dR^2 - dL^2,
\end{equation}
where $L=\psi_0$ is a constant and $H_c(t)=\dot a/a$ is the classical
Hubble parameter.
The new variables has physical meaning, because
$t$ is the cosmic time and $(R,L)$ are spatial variables.
Furthermore $a(t)$ is the scale factor of the
universe and describes its 3D euclidean (spatial) volume.
Hence the effective 4D metric
is a spatially (3D) flat FRW one
\begin{equation}\label{frw}
dS^2 \rightarrow ds^2 = dt^2 - e^{2\int H_c(t) dt} dR^2,
\end{equation}
and has a scalar curvature $^{(4)}{\cal R} = 6(\dot H_c + 2 H^2_c)$. The
metric (\ref{frw}) has a metric tensor with components $g_{\mu\nu}$
($\mu,\nu$ take the values $0,1,2,3$).
The determinant of this tensor is $^{(4)}g=(a/a_0)^6$.
Furthermore, the metric (\ref{frw}) describes globally a isotropic
and homogeneous universe.

Now we can make the same treatment to the density Lagrangian
(\ref{1'}) and the differential equation
(\ref{df}). Using the transformations (\ref{*}) we obtain
\begin{eqnarray}
&& ^{(4)}{\cal L}\left[\varphi(\vec{R},t), \varphi_{,\mu}(\vec{R},t)\right]
=\frac{1}{2} g^{\mu\nu} \varphi_{,\mu} \varphi_{,\nu} -
\frac{1}{2} \left[(R H_c)^2  -
\frac{a^2_0}{a^2} \right] \  \left(\nabla_R \varphi\right)^2, \label{aa}\\
&& \ddot\varphi + 3 H_c\dot\varphi -\frac{a^2_0}{a^2} \nabla^2_R \varphi
+
\left[\left(4\frac{H^3_c}{\dot H_c} - 3\frac{\dot H_c}{H_c} -
3\frac{H^5_c}{\dot H^2_c}\right) \dot\varphi +
\left(\frac{a^2_0}{a^2} - H^2_c R^2\right)\nabla^2_R\varphi\right]=0.
\label{bb}
\end{eqnarray}
Hence, with this representation the effective scalar 4D potential
$V(\varphi)$ and its derivative with respect
to $\varphi(\vec{R},t)$ are
\begin{eqnarray}
&& V(\varphi) \equiv \frac{1}{2}\left[ (R H_c)^2
- \left(\frac{a_0}{a}\right)^2 \right] \left(\nabla_R\varphi\right)^2,
\label{au} \\
&&
V'(\varphi) \equiv 
\left(4\frac{H^3_c}{\dot H_c} - 3\frac{\dot H_c}{H_c} -
3\frac{H^5_c}{\dot H^2_c}\right) \dot\varphi +
\left(\frac{a^2_0}{a^2} - H^2_c R^2\right)\nabla^2_R\varphi.\label{a1}
\end{eqnarray}
The equations (\ref{aa}) and (\ref{bb}) describe the dynamics of the inflaton field
$\varphi(\vec{R},t)$ in a metric (\ref{frw}) with a Lagrangian
\begin{equation}\label{l4}
^{(4)}{\cal L}[\varphi(\vec{R},t),\varphi_{,A}(\vec{R},t)] =
-\sqrt{\left|\frac{^{(4)}g}{^{(4)}g_0}\right|}
\left[\frac{1}{2} g^{\mu\nu} \varphi_{,\mu}\varphi_{,\nu}
+V(\varphi)\right],
\end{equation}
where $\left|^{(4)}g_0\right|=1$.
In the new representation $(R,t,L)$, we obtain the following new
velocities $ \hat U^A ={\partial \hat x^A \over \partial x^B} U^B$
\begin{equation} \label{10}
U^t=\frac{2u(t)}{\sqrt{u^2(t)-1}}, \qquad
U^R=-\frac{2r}{\sqrt{u^2(t)-1}}, \qquad U^L=0,
\end{equation}
where
the old velocities $U^B$ are $U^N$, $U^r=0$ and $U^{\psi}$.
Furthermore, the velocities $\hat U^B$ complies with the
constraint condition
\begin{equation}\label{con}
\hat g_{AB} \hat U^A \hat U^B =1.
\end{equation}
The important fact here is that the new frame give us an effective
spatially flat FRW metric embedded in a 5D manifold where
the initial value of the
fifth coordinate $L_0=\psi_0=1/H_0$ is the primordial Hubble horizon,
which emerges naturally as a constant in this representation.

The solution $N={\rm arctanh}[1/u(t)]$ corresponds to a power-law
expanding universe with time dependent power $p(t)$
for a scale factor $a \sim t^{p(t)}$. Since $H_c(t) = \dot a/a$, the
resulting Hubble parameter is
\begin{equation}\label{h}
H_c(t)=\dot p {\rm ln}(t/t_0) +p(t)/t,
\end{equation}
where $t_0$ is the time for which inflation ends.
The function $u$ written as a function of time is
\begin{equation}
u(t) = -\frac{H^2_c}{\dot H_c},
\end{equation}
where the overdot represents the derivative with respect to the time.
In this frame, the 4D energy 
density $\rho$ and the pressure ${\rm p}$ are\cite{PLB}
\begin{eqnarray}
&& 8 \pi G \rho = 3 H^2_c,\\
&& 8\pi G {\rm p} = -(3H^2_c + 2 \dot H_c).
\end{eqnarray}
Furthermore, note that the condition (\ref{con}) implies that
$|u(t)|=\sqrt{{4 r^2 (a/a_0)^2 -1 \over 3}} >1$, where $r$ is a
constant. Moreover, the function $u(t)$ can be related
to the deceleration parameter $q(t)=-\ddot a a/\dot a^2$: $u(t)=1/[1+q(t)]$,
such that for inflationary models the required condition
$|q(t)| \simeq 1$ (but with negative $q$), is fulfilled for
$r =R(t) H_c(t) \gg 1$.
In other words, it means that
the effective 4D background metric (\ref{frw})
is only valid on super Hubble scales: $R \gg 1/H_c$, in agreement
with the expected for a background metric.
Note that the function $1/u(t)=-\dot H/H^2 \ll 1$
give us the slow-roll parameter\cite{liddle}
during inflation\cite{BCMS}. So, the feasible values
for the constant $r$ during inflation being given only from geometrical
arguments. This is an important prediction of the model here developed.

On the other hand, $V(\varphi)$ and $V'(\varphi)$ can
be written as a function of the old coordinates $(N,r,\psi)$
in the comoving frame $U^r=0$
\begin{eqnarray}
&& V(\varphi) \equiv \frac{1}{2} \left[r^2 - e^{-2N}\right] \frac{1}{r^2}
\left(\frac{1}{\stackrel{\star}{\psi}} \stackrel{\star}{\varphi}\right)^2,
\label{v1} \\
&& V'(\varphi) \equiv
\left(3 \frac{\stackrel{\star}{\psi}}{\psi^3}
-\frac{4}{\psi\stackrel{\star}{\psi}}
-\frac{3}{\stackrel{\star}{\psi}^2}\right)
\stackrel{\star}{\varphi}+
\left[\left(\frac{a_0}{a}\frac{1}{r}\right)^2-1\right] \frac{
\partial^2\varphi}{\partial \psi^2}.\label{v'}
\end{eqnarray}
Here, the overstar denotes
the derivative with respect to $N$.
Note that $\Delta N$ is the number of e-folds of the universe. To inflation
solves the horizon/flatness problems it is required that $\Delta N\ge 60$
at the end of inflation.

At this point we can introduce the 4D
Hamiltonian ${\cal H}=\pi^0 \dot\varphi-
^{(4)}{\rm L}$, where the 4D Lagrangian is
$^{(4)} {\rm L}(\varphi,\varphi_{,\mu})
= \sqrt{{|^{(4)}g|
\over |^{(4)}g_0| }} \  {^{(4)}{\cal L}}(\varphi,\varphi_{,\mu})$
[see eq. (\ref{l4})]:
\begin{equation}
{\cal H} = \frac{1}{2} \frac{a^3}{a^3_0}
\left[ \dot\varphi^2 + \frac{a^2_0}{a^2} \left(\nabla\varphi\right)^2
+2V(\varphi)\right].
\end{equation}
Hence, we can define the energy density operator $\rho$ such that
${\cal H} = {a^3 \over a^3_0} \rho$. Hence
\begin{equation}
\rho = \frac{1}{2} 
\left[ \dot\varphi^2 + \frac{a^2_0}{a^2} \left(\nabla\varphi\right)^2
+2V(\varphi)\right].
\end{equation}
The 4D expectation value of the Einstein equation
$H^2 = {8\pi G \over 3} \rho$ on the 4D FRW metric (\ref{frw}), will be
\begin{equation}
\left<H^2\right> = \frac{4\pi G}{3} \left< \dot\varphi^2 +
\frac{a^2_0}{a^2} \left(\nabla\varphi\right)^2 + 2 V(\varphi)\right>,
\end{equation}
where $G$ is the gravitational constant.
We can make a semiclassical treatment\cite{BCMS,NPB} for the quantum field
$\varphi$, such that $<\varphi(\vec{R},t)> = \phi_c(t)$:
\begin{equation}
\varphi(\vec{R},t) = \phi_c(t) + \phi(\vec{R},t),
\end{equation}
where $<\phi>=0$. Furthermore, we impose that $<\dot\phi>=0$.
With this approach the classical dynamics on the background 4D FRW
metric (\ref{frw}) is well described by the equations
\begin{eqnarray}
&& \ddot\phi_c + 3 \frac{\dot a}{a} \dot\phi_c + V'(\phi_c)=0, \label{ua} \\
&& H^2_c = \frac{8\pi G}{3} \left(\frac{\dot\phi^2_c}{2} + V(\phi_c)
\right).\label{hc}
\end{eqnarray}
Since $\dot\phi_c=-{H'_c \over 4\pi G}$, from eq. (\ref{hc}) we obtain
the classical scalar potential $V(\phi_c)$ as a function of the
classical Hubble parameter $H_c$
\begin{displaymath}
V(\phi_c) = \frac{3 M^2_p}{8\pi} \left[ H^2_c - \frac{M^2_p}{12\pi}
\left(H'_c\right)^2 \right],
\end{displaymath}
where $M_p=G^{-1/2}$ is the Planckian mass.
The quantum dynamics is described by
\begin{eqnarray}
&& \ddot\phi + 3 \frac{\dot a}{a} - \frac{a^2_0}{a^2} \nabla^2\phi +
\sum_{n=1} \frac{1}{n!} V^{(n+1)}(\phi_c) \phi^n =0, \label{apr} \\
&& \left< H^2 \right> =  H^2_c + \frac{8\pi G}{3}
\left< \frac{\dot\phi}{2} + \frac{a^2_0}{2 a^2} (\nabla\phi)^2 +
\sum_{n=1} \frac{1}{n!} V^{(n)}(\phi_c) \phi^n \right>.\label{fried}
\end{eqnarray}
On cosmological scales, the quantum fluctuations are small, so that
a linear approximation (i.e., $n=1$) is sufficient to make a
realistic description for the evolution of $\phi$. Furthermore,
the second term in (\ref{fried}) is negligible when $\phi$ is considered
spatially very homogeneous. However, such that term could be very
important on sub Hubble scales\cite{Nambu}. With the aim to make a description
of the dynamics on cosmological scales, we shall consider this term
as null. For this reason we shall take
${\dot a^2\over a^2} = <H^2> \simeq H^2_c$.
Once done the linear approximation for the semiclassical treatment
we can make the identification of the squared mass for the
inflaton field $m^2 = V''(\phi_c)$. Hence, after make a linear expansion
for $V'(\varphi)$ in eq. (\ref{v'}), we obtain
\begin{eqnarray}
&& V'(\phi_c) =
\left(4\frac{H^3_c}{\dot H_c} -
3\frac{\dot H_c}{H_c} - 3 \frac{H^5_c}{\dot H^2_c}\right)
\dot\phi_c,\label{ua1} \\
&& m^2 \phi \equiv
\left(4 \frac{H^3_c}{\dot H_c} - 3 \frac{\dot H_c}{H_c} - 
3\frac{H^5_c}{\dot H^2_c} \right) \frac{\partial \phi}{\partial t} +
\left( \frac{a^2_0}{a^2} - H^2_c R^2 \right) \nabla^2_R\phi.\label{apr1}
\end{eqnarray}
Taking into account the expressions (\ref{ua}) with (\ref{ua1}) and
(\ref{apr}) with (\ref{apr1}), we obtain the dynamics for $\phi_c$ and $\phi$.
Hence, the equations $\ddot\phi_c+ 3H_c \dot\phi_c + V'(\phi_c)=0$
and $\ddot\phi +   3H_c\dot\phi_c - (a/a_0)^2\nabla^2_R \phi
+ V''(\phi_c)\phi=0$ now take the form
\begin{eqnarray}
&& \ddot\phi_c + \left[3H_c+ f(t)\right]\dot\phi_c =0, \\
&&  \ddot\phi + \left[3H_c(t) + f(t)\right]\dot\phi -
H^2_cR^2\nabla^2_R\phi =0, \label{qf}
\end{eqnarray}
where
\begin{equation}\label{f}
f(t)=\left(4\frac{H^3_c}{\dot H_c} -
3\frac{\dot H_c}{H_c} - 3 \frac{H^5_c}{\dot H^2_c}\right).
\end{equation}
Moreover, $V'(\phi_c)$ and $V''(\phi_c)\phi$ can be written
in terms of the metric (\ref{6}), for a comoving observer (with
$U^r=0$)
\begin{eqnarray}
&& V'(\phi_c) =
\left[ 3 \frac{\stackrel{\star}{\psi}}{\psi^3}
-\frac{4}{\psi\stackrel{\star}{\psi}}
- \frac{3}{\stackrel{\star}{\psi}^2} \right]
\stackrel{\star}{\phi_c}, \\
&& m^2 \phi \equiv
\left[ 3 \frac{\stackrel{\star}{\psi}}{\psi^3}
-\frac{4}{\psi\stackrel{\star}{\psi}}
- \frac{3}{\stackrel{\star}{\psi}^2} \right]
\stackrel{\star}{\phi}
+ \left[ \left(\frac{a_0}{a} \frac{1}{r} \right)^2 -1\right]
\frac{\partial^2\phi}{\partial\psi^2}.
\end{eqnarray}
Note that in this representation
{\rm a)} the squared mass is a differential operator that acts on
the quantum fluctuations $\phi$, and
{\rm b)} the 4D
potential and its derivatives with respect to $\varphi(\vec{R},t)$
are consequence of
the evolution of $\psi(N)$. In other words the nonzero curvature of the
4D potential is induced by the geodesic
evolution of the fifth coordinate
for an observer in a comoving frame with $U^r=0$.
So, the inflationary dynamics, which is described by the inflaton
field, should be determined by the evolution of the space-like
fifth coordinate on a foliation of the 5D metric (\ref{6})
where $r$ is a constant. This is the main
result of this letter.

This formalism could take important consequences in the early
universe. During the inflationary epoch, the slow-roll condition
$\gamma(t) = -\dot H_c/H^2_c \ll 1$ holds.
Since $u(t)=1/\gamma(t)$,
we obtain that $u\gg 1$. This assures that all the velocities
in $U^A$ in (\ref{vel}) and $\hat U^A$ in (\ref{10}) to be real, and
imposes the condition $r \gg 1$\cite{MB}.
Furthermore the equation of state can be written in terms
of the function $u(t)$
\begin{displaymath}
\left<{\rm p}\right> = - \left[1-\frac{2}{3 u(t)}\right] \left<\rho\right>,
\end{displaymath}
which, since $u \gg 1$ during inflation, complies with the required
condition for this stage: $\left<{\rm p}\right> \simeq - \left<\rho\right>$.
Moreover, speaking in terms of the effective 4D FRW metric (\ref{frw}),
the geodesic
evolution of the fifth coordinate give us the Hubble horizon $\psi(t)=
1/H(t)$ and the resulting fifth (constant) coordinate 
$L=\psi_0$ is given by the primordial Hubble horizon:
$L=1/H_c(t_0)$.

We can define the {\it redefined quantum fluctuations} $\chi(\vec{R},t)=
e^{1/2 \int [3H_c(t)+f(t)]dt} \phi$, so that the equation of motion
for $\chi$ yields
\begin{equation}
\ddot\chi - \left[H^2_c R^2 \nabla^2_R + \frac{1}{4}
\left(3H_c + f(t)\right)^2 + \frac{1}{2} \left(3\dot H_c + \dot f(t)\right)
\right]\chi =0,
\end{equation}
so that the modes $\chi_k(\vec{R},t)$ of the field $\chi$ complies
the differential equation
\begin{equation}\label{eq1}
\ddot\chi_k + H^2_c R^2 \left(k^2 - k^2_0(t)\right) \chi_k=0,
\end{equation}
with
\begin{equation}\label{k0}
k^2_0(t) = \frac{1}{R^2H^2_c} \left[\frac{1}{4} \left(3H_c + f(t)\right)^2
+ \frac{1}{2} \left(3\dot H_c + \dot f(t) \right)\right],
\end{equation}
where $f(t)$ is a function of the classical Hubble parameter [see eq.
(\ref{f})]. Hence, all the dynamics of the quatum fluctuations being
described only by the classical Hubble parameter $H_c=\dot a/a$.

To ilustrate the formalism we can study an example where
$\psi(N)=-1/(\alpha N)$, so that $H_c(N)=-\alpha N$. This implies that
the classical
Hubble parameter (written as a function of time)
is given by $H_c(t)=H_0 e^{\alpha \Delta t}$. At the end of inflation
$\alpha \Delta t \ll 1$, so that $H_c(t) \simeq H_0 (1+\alpha \Delta t)$ and
$3H_c(t) + f(t) \simeq 3H_0 (1+\alpha \Delta t)
+ 3\alpha -(4 H^2_0/\alpha) (1+2\alpha \Delta t)
-(3 H^3_0/\alpha^2) (1+3\alpha \Delta t)$,
where $\Delta t = t_0 -t$ and $t_0$ is the time for which inflation ends.
At the end of inflation it is sufficient to make a
$\Delta t$-first order expansion for $k^2_0$, so that it can be
approximated to
\begin{equation}
k^2_0(t) = \frac{1}{r^2} \left(A-B t\right).
\end{equation}
With this approximation, the general solution
for the modes $\chi_k(t)$ is
\begin{equation}
\chi_k(t) = C_1 {\rm Ai}\left[x(t)\right] + C_2 {\rm Bi}\left[x(t)\right],
\end{equation}
where
${\rm Ai}\left[x(t)\right]$ and ${\rm Bi}\left[x(t)\right]$
are the Airy functions with argument $x(t)$. Furthermore,
($C_1$,$C_2$) are some constants and
\begin{eqnarray}
A& =& \frac{1}{4} \left(3 H_0 - 3 \frac{H^3_0}{\alpha^2} + \alpha -
3\frac{H^2_0}{\alpha}\right)^2 + \frac{1}{2} \left(8 H^2_0
- 9\frac{H^3_0}{\alpha} -\alpha H_0\right) \nonumber \\
& - & \frac{1}{2}
\left(3 H_0 - 3 \frac{H^3_0}{\alpha^2} + 3\alpha -
8\frac{H^2_0}{\alpha}\right) \left(8H^2_0 + 9 \frac{H^3_0}{\alpha} -
3H_0 \alpha\right) t_0, \\
B  & =&  \frac{1}{2} \left(3 H_0 - 3 \frac{H^3_0}{\alpha^2} + 3\alpha
-8\frac{H^2_0}{\alpha}\right)\left(3H_0 \alpha -8 H^2_0 -
9 \frac{H^3_0}{\alpha}\right), \\
x(t) & = & \frac{\left[(A-k^2)-B t \right]}{b}
\left(\frac{b}{r^2}\right)^{1/3}.
\end{eqnarray}
Note that in this example $H_0$ denotes the value of the Hubble parameter
at the end of inflation.
On cosmological scales (i.e., for $k^2 \ll A - Bt$), the
solution for $\chi_k$ is unstable.
However in the UV sector (i.e., for $k^2 \gg A-B t$),
the modes oscillate. This behavior is well described by the
function ${\rm Bi}[x(t)]$, so that we shall take
$C_1=0$. Hence, at the end of inflation the modes $\chi_k$ will
be
\begin{equation}\label{ch}
\chi_k(t) = C_2 \  {\rm Bi}[x(t)].
\end{equation}
Since the modes of the quantum fluctuations $\phi$
are $\phi_k=e^{-1/2\int [3H_c+f(t)]dt} \chi_k $, 
the squared fluctuations
are given by
\begin{equation}\label{sf}
\left<\phi^2\right> \simeq \frac{1}{2\pi^2}
e^{-\left[3(H_0+\alpha)-4\frac{H^2_0}{\alpha}
-3\frac{H^3_0}{\alpha^2}\right]t}
{\Large\int} dk \  k^2 \left|\chi^2_k\right|,
\end{equation}
where the modes $\chi_k$ are given by eq. (\ref{ch}).
Furthermore the density fluctuations at the end of inflation
can be estimated by the expression
\begin{equation}
\frac{\delta\rho}{\rho} \sim  \frac{H^2_0}{\dot\phi_c}
\sim 2 \pi^{1/2} \frac{H^{3/2}_0 }{M_p \alpha^{1/2}},
\end{equation}
which are of the order of $10^{-5}$ for $H_0 \sim 10^{-5} \  M_p$
and $\alpha \sim 10^{-5} \  M_p$.
In our case, the spectral index $n_s$ being given by
$n_s -1=-{6\over u(t)}$. During inflation $u \gg 1$, so that
$|n_s-1| \ll 1$. Hence, during inflation the spectrum approaches
very well with a Harrison - Zeldovich one.
A more exaustive treatment for density fluctuations go beyond
the scope of this letter.

In this letter we have studied a single scalar field inflationary
model which emerges from a 5D apparent vacuum described by
a flat 5D metric with coordinates
($N$,$r$,$\psi$) and a Lagrangian for a free scalar field.
The interesting is that the scalar potential $V(\varphi)$
appears
in the 3D comoving frame characterized by $U^r=0$ [see eq. (\ref{v1})].
A further transformation to physical coordinates
$t=\int \psi(N) dN$, $R=r\psi$ and $L=\psi e^{-\int dN/u(N)}$ give
us the possibility to describe the system in a effective 4D
(but 3D spatially flat) FRW metric.
Such that metric is view as a particular frame (characterized with $U^L=0$)
of the 5D metric (\ref{fr}),
where the potential
$V(\varphi)$ is represented as the differential operator
(\ref{au}). In other words, the potential,
which assume different representations in different
frames, has a geometrical origin.
Moreover, the mass of the inflaton field
appears in the frame $U^L=0$ as a differential operator applied
to que quantum fluctuations $\phi(\vec R,t)$. Hence, for the
semiclassical (and linear on $\phi$) treatment here developed,
$m^2\phi(\vec R,t)$
is a local operator with nonzero expectation value.
At this point we must to exalt this result, because
a particular frame in physics is intrinsically related
to an observer (or experimental result).
Note that in the potential (\ref{au}) the KK modes are excluded.
These modes should be related to a spin-2 graviton that appears
in the KK theory when the electromagnetic effects are included.
This is not our case. In this letter we are excluded the electromagnetics
fields in the metric because the electromagnetics effects should be non
important on cosmological scales.

To conclude, this formalism could be generalized to other
inflationary models with many scalar fields\cite{cas}.
Deflationay models could be examined using a line element
$S(N)=N$ (we emphasize that here we have used $S(N)=-N$ which
describes expanding universes).
Moreover, models with nonzero cosmological parameters could be
studied. The formalism also could be successfully
extended to the study of topological defects\cite{vil}
by using a line element $S\equiv S(r)$ on some frame
$U^N=0$, rather the here used $S\equiv S(N)$.
\vskip .2cm
\centerline{\bf{Acknowledgements}}
\vskip .2cm
JEMA acknowledges CONACyT and IFM of UMSNH (M\'exico)
for financial support.
MB acknowledges CONICET, AGENCIA 
and UNMdP (Argentina)
for financial support.\\


\begin{thebibliography}{99}
\bibitem{*} V. A. Rubakov and M. E. Shaposhnikov, Phys. Lett. {\bf B125}, 136
(1983); I. Antoniadis, N. Arkani-Hamed, S. Dimopoulos and
G. Dvali, Phys. Lett. {\bf B436}, 257 (1998);
G. Shiu and S. H. Tye, Phys. Rev. {\bf D58}, 106007 (1998);
R. Sundrum, Phys. Rev. {\bf D59}, 085009 (1999);
Z. Kakushadze and S. H. H. Tye, Nucl. Phys. {\bf B548}, 180 (1999);
L. Randall and R. Sundrum, Phys. Rev. Lett. {\bf 83}, 4690 (1999).
\bibitem{a} V. A. Rubakov and M. E. Shaposhnikov, Phys. Lett. {\bf 125}, 139
(1983).
\bibitem{b} M. J. Duff, B. E. W. Nilsson, C. N. Pope, Phys. Rep. {\bf 130},
1 (1986);
I. Antoniadis, Phys. Lett. {\bf 246}, 377 (1990);
J. Mourad, Class. Quant. Grav. {\bf 10}, 2157 (1997);
Hong-Ya Liu, P. S. Wesson, Mod. Phys. Lett. {\bf A13}, 2689 (1998);
A. Mazumdar, Phys. Lett. {\bf B469}, 55 (1999);
R. Brandenberger, D. A. Easson, A. Mazumdar, Phys. Rev. {\bf D69},
083502 (2204).
\bibitem{a1} M. Gell-Mann and B. Zwiebach, Nucl. Phys. {\bf 260}, 56 (1985).
\bibitem{Guth}
A. A. Starobinsky,
JETP Lett. {\bf 30}, 682 (1979); Phys. Lett. {\bf B91},
99 (1980);
A. Guth, Phys. Rev. {\bf D23}, 347 (1981);
A. Guth, E. J. Weinberg, Nucl. Phys. {\bf B212}, 321 (1983);
A. Albrecht and P. J. Steinhardt, Phys. Rev. Lett. {\bf 48}, 1220 (1982);
A. D. Linde, Phys. Lett. {\bf 129}, 177 (1983);
For a review of inflation the reader can see A. D.
Linde, {\em Particle Physics and Inflationary Cosmology}, Harwood,
Chur, Switzerland, 1990; A. D. Linde, Phys. Rep. {\bf 333},
575 (2000).
\bibitem{wesson} J. M. Overduin, P. S. Wesson, Phys. Rep. {\bf 283},
303 (1997).
\bibitem{PLB} D. S. Ledesma and M. Bellini, Phys. Lett. {\bf B581}, 1 (2004).
\bibitem{MB} E. Madriz Aguilar and M. Bellini,
{\em Origin of FRW cosmology in slow roll inflation
from noncompact Kaluza-Klein theory},
E-print: gr-qc/0402108.
\bibitem{liddle} E. J. Copeland, E. W. Kolb, A. R. Liddle and J. E. Lidsey,
Phys. Rev. {\bf D48}, 2529 (1993).
\bibitem{BCMS} M. Bellini, H. Casini, R. Montemayor and P. Sisterna,
Phys. Rev. {\bf D54}, 7172 (1996).
\bibitem{NPB} M. Bellini, Nucl. Phys. {\bf B604}, 441 (2001).
\bibitem{Nambu} Y. Nambu, Phys. Rev. {\bf D65}, 104013 (2002).
\bibitem{cas} M. Castagnino, H. Giacomini and L. Lara,
Phys. Rev. {\bf D65}, 023509 (2002).
\bibitem{vil} R. Basu and A. Vilenkin, Phys. Rev. {\bf D50}, 7150 (1994);
V. Berezinsky, P. Blasi and A. Vilenkin, Phys. Rev. {\bf D58},
103515 (1998).



\end{thebibliography}
\end{document}